\documentclass[lettersize,journal]{IEEEtran}
\usepackage{amsmath,amsfonts}
\usepackage{algorithmic}
\usepackage{array}
\usepackage{booktabs}
\usepackage[caption=false,font=normalsize,labelfont=sf,textfont=sf]{subfig}
\usepackage{textcomp}
\usepackage{stfloats}
\usepackage{url}
\usepackage{verbatim}
\usepackage{graphicx}
\def\BibTeX{{\rm B\kern-.05em{\sc i\kern-.025em b}\kern-.08em
    T\kern-.1667em\lower.7ex\hbox{E}\kern-.125emX}}
\usepackage{balance}
\begin{document}
\title{A Unified Fully Reconfigurable Architecture for Wireless Powered Communication Networks}
\author{Bingxin~Zhang,~\IEEEmembership{Member,~IEEE,}
	Yizhe~Zhao,~\IEEEmembership{Member,~IEEE,}
	and Kun~Yang,~\IEEEmembership{Fellow,~IEEE}
	\thanks{Bingxin Zhang, Kun Yang  are with the State Key Laboratory of Novel Software Technology, Nanjing University, Nanjing, 210008, China, Institute of Intelligent Networks and Communications (NINE), Collaborative Innovation Center of Novel Software Technology and Industrialization, and School of Intelligent Software and Engineering, Nanjing University (Suzhou Campus), Suzhou, 215163, China (email: bxzhang@nju.edu.cn; kunyang@nju.edu.cn).}
	\thanks{Yizhe Zhao is  with the School of Information and Communication Engineering, University of Electronic Science and Technology of China, Chengdu 611731, China (e-mail: yzzhao@uestc.edu.cn).}
}

\maketitle

\begin{abstract}
Wireless powered communication networks (WPCNs) are a key enabler for sustainable Internet of Things (IoT) systems, yet their practical performance is constrained by inefficient wireless energy transfer, limited spatial adaptability, and fragile uplink connectivity in blockage-prone and dynamic environments. Emerging reconfigurable antenna technologies, including pinching antenna systems (PASSs), fluid antenna systems (FASs), movable antennas (MAs), and reconfigurable intelligent surfaces (RISs), provide new opportunities to overcome these limitations, but have mostly been studied separately. In this article, we propose a unified architecture for fully reconfigurable WPCNs by jointly integrating PASS-enabled power beacons, FAS-based IoT devices, MA-assisted base stations, and RIS-aided propagation environments. The proposed framework enables end-to-end reconfigurability across downlink energy transfer, device-side spatial adaptation, base-station reception, and uplink information transmission. We further discuss the integration motivation, system architecture, design and optimization framework, illustrative performance evaluation, implementation tradeoffs, and major practical challenges. This article provides a new perspective for designing next-generation fully reconfigurable WPCNs.
\end{abstract}


\section{Introduction}

\IEEEPARstart{F}{uture} Internet of Things (IoT) networks are expected to support massive low-power devices in smart factories, intelligent buildings, environmental monitoring systems, transportation networks, and low-altitude platforms \cite{SBSAH}. For these devices, sustainable operation remains a fundamental challenge, since frequent battery replacement or manual recharging is costly and often infeasible in large-scale or hard-to-access deployments. Wireless powered communication networks (WPCNs), where IoT devices harvest radio-frequency (RF) energy and use the harvested energy for uplink information transmission, offer a promising paradigm for energy-sustainable IoT networks \cite{DNDIK}.

A key limitation of conventional WPCNs lies in their spatially static architecture. In practical deployments, irregular device distributions, blockage, and propagation loss may weaken downlink wireless energy transfer (WET), while uplink wireless information transmission (WIT) is further limited by the harvested energy. Although joint beamforming and resource allocation have been widely studied to improve WPCN performance \cite{HPYL}, most existing designs still operate under fixed network geometries, antenna configurations, and propagation environments. Therefore, their adaptability mainly comes from parameter optimization, rather than from reshaping the spatial structure of the energy-information transfer process. Since WET and WIT are tightly coupled, future WPCNs need to move toward a system-level architecture with end-to-end reconfigurability.

A new generation of flexible antenna and surface technologies is changing this picture \cite{YZ}. Pinching antenna systems (PASSs) \cite{YLZW}, fluid antenna systems (FASs) \cite{HYYZ}, movable antennas (MAs) \cite{BNea}, and reconfigurable intelligent surfaces (RISs) \cite{CP} provide different ways to make wireless networks spatially adaptive, such as adjusting pinching points, antenna states, antenna positions, and propagation paths. Recent studies have begun to explore reconfigurable WPCNs enabled by FASs, PASSs, or RISs \cite{XLYZ,YLHX,KT}, while pairwise combinations of reconfigurable antenna and surface technologies have also been investigated for wireless performance enhancement \cite{YGQW,JY,CHYL,XLYZ2}. These efforts indicate a clear trend toward more flexible wireless infrastructures. For WPCNs, however, the key issue is not merely to introduce more reconfigurable components, but to coordinate them under a unified architecture so that energy transfer and information transmission can be jointly adapted to device distribution, channel variation, and environmental blockage.

This integration is motivated by three main considerations. First, PASS-enabled PBs can shorten the effective WET distance and improve the harvested-energy budget, thereby extending self-sustainable service coverage. Second, FASs, MAs, and RISs can enhance device-side adaptation, BS-side reception, and uplink propagation control, making the WIT process more robust against fading and blockage. Third, since WET and WIT are tightly coupled through the harvested energy, jointly coordinating PASS, FAS, MA, and RIS can improve the end-to-end energy-information efficiency, rather than optimizing the energy or information link separately. Such an architecture is particularly attractive for battery-free smart manufacturing, environmental sensing, and smart city infrastructure monitoring, where low-power devices often suffer from non-uniform energy supply, dynamic blockage, and unreliable uplink connectivity.

Motivated by these observations, this article introduces fully reconfigurable WPCNs, where heterogeneous reconfigurable components are coordinated so that energy transfer and information transmission can be adapted as an end-to-end process. The main contributions of this article are summarized as follows.

\begin{itemize}
	\item We present a unified architecture for fully reconfigurable WPCNs by integrating PASS-enabled PBs, FAS-based IoT devices, MA-assisted BSs, and RIS-aided propagation environments, thereby enabling end-to-end spatial reconfigurability across the coupled WET-WIT process.
	
	\item We clarify the motivation and potential benefits of integrating heterogeneous reconfigurable components, showing how coordinated PASS, FAS, MA, and RIS reconfiguration can extend self-sustainable service coverage, enhance uplink robustness, and improve energy-information efficiency.
	
	\item We develop a general design and optimization framework for fully reconfigurable WPCNs, covering indicator determination, optimization problem formulation, and optimization problem resolution under the coupled WET-WIT process.
	
	\item We provide illustrative performance evaluation and implementation tradeoff analysis to demonstrate the potential gain and practical cost of fully reconfigurable WPCNs, followed by discussions on key challenges and future research directions.
\end{itemize}

\begin{figure*}[!t]
	\centering
	\includegraphics[width=0.68 \textwidth]{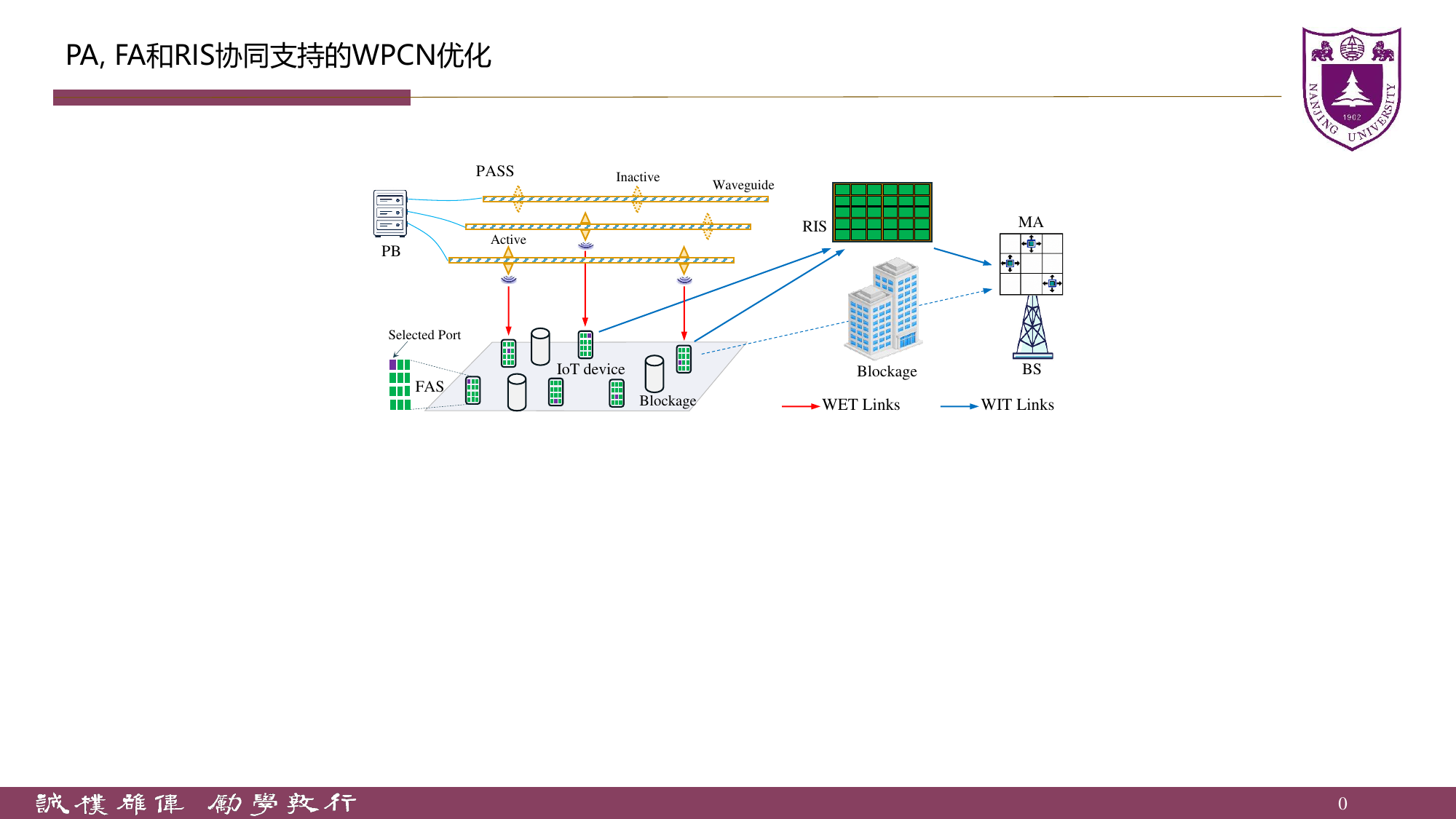}
	\caption{The unified architecture of fully reconfigurable WPCNs.
	}
	\label{fig:architecture}
\end{figure*}

\begin{table*}[!t]
	\renewcommand{\arraystretch}{1.15}
	\caption{Comparison of Reconfigurable Components in Fully Reconfigurable WPCNs}
	\label{tab:components}
	\centering
	\scriptsize
	\setlength{\tabcolsep}{2pt}
	\begin{tabular*}{0.82\textwidth}{@{\extracolsep{\fill}}p{0.10\textwidth}p{0.24\textwidth}p{0.36\textwidth}p{0.12\textwidth}}
		\toprule
		Component &
		Key Reconfiguration Domain &
		Role in Fully Reconfigurable WPCNs &
		Network Stage \\
		\midrule
		PASS &
		Pinching-point activation &
		PB-side short-distance LoS WET link establishment &
		WET \\
		FAS &
		Antenna port selection &
		Device-side local channel diversity exploitation &
		WET / WIT \\
		MA &
		Antenna position optimization &
		BS-side receive channel matching &
		WIT \\
		RIS &
		Reflection-coefficient control &
		Environment-side propagation reconfiguration for WIT &
		WIT \\
		\bottomrule
	\end{tabular*}
\end{table*}

\section{Unified Architecture for Fully Reconfigurable WPCNs}

Fully reconfigurable WPCNs are designed to adapt the energy transmitter, IoT devices, information receiver, and wireless environment to changing service and channel conditions. This section presents a representative architecture to clarify their functional roles and coordinated energy-information transfer process.

\subsection{Structural Composition of Fully Reconfigurable WPCNs}

As shown in Fig. \ref{fig:architecture}, the main components of fully reconfigurable WPCNs include the PASS-enabled PB, FAS-based IoT devices, MA-assisted BS, and RIS-aided propagation environment. These components introduce heterogeneous forms of spatial reconfigurability across different segments of the network, enabling adaptive control of WET and WIT.

Table \ref{tab:components} further summarizes the functional roles and reconfiguration capabilities of these components in a unified manner, highlighting their distinct contributions to energy transfer, signal reception, device-side adaptation, and environment-level propagation control.

\textit{1) PASS-enabled PB:} The PB provides downlink RF energy supply. By connecting the PB to dielectric waveguides, pinching antennas can be deployed along the waveguides to form flexible pinching points across the target region, allowing the energy source to better adapt to device locations and propagation conditions. Compared with conventional fixed PBs, this architecture introduces additional spatial reconfigurability for WET.

\textit{2) FAS-based IoT Device:} IoT devices harvest energy in the downlink WET phase and transmit information in the uplink WIT phase. With FASs, each device can switch among multiple antenna states or ports within a compact space, enabling device-side adaptation to local channel variations. This is particularly beneficial for low-power devices with limited energy and hardware resources.

\textit{3) MA-assisted BS:} The BS receives uplink signals from IoT devices and also serves as a coordination center for the system. By employing MAs, it can adjust antenna positions within a predefined region to enhance the effective uplink channel and improve reception reliability. In addition, the BS collects channel state information (CSI), energy-state information, and device status to support system-level coordination.

\textit{4) RIS-aided Propagation Environment:} RISs are deployed between IoT devices and the BS to establish controllable propagation paths for uplink transmission. By tuning the reflection coefficients of RIS elements, the uplink channel can be reshaped to mitigate blockage, enhance useful signal components, and improve WIT reliability. As a result, the propagation environment becomes a controllable part of the fully reconfigurable WPCN architecture.

\subsection{Energy-Information Transfer Process and Coordination}

\begin{figure*}[!t]
	\centering
	\includegraphics[width=0.96 \textwidth]{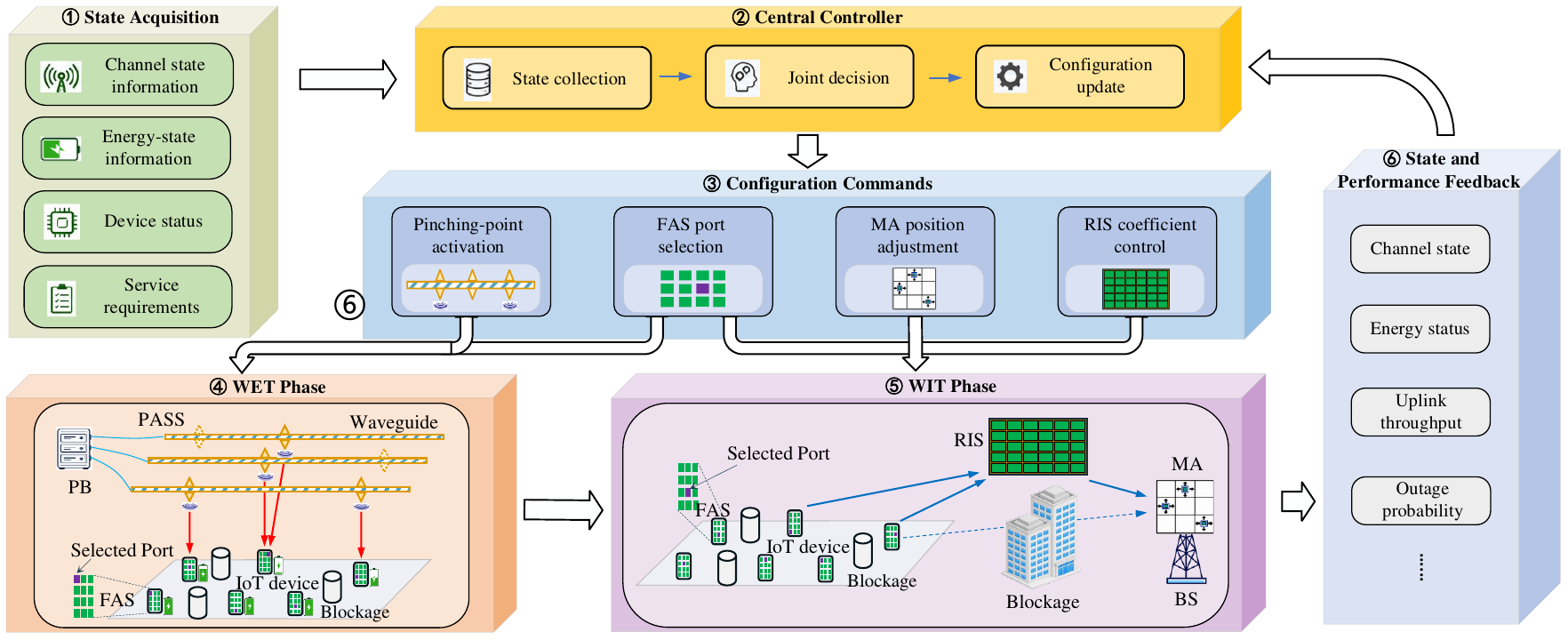}
	\caption{Workflow of fully reconfigurable WPCNs with coordinated WET-WIT reconfiguration.}
	\label{fig:workflow}
\end{figure*}

The operation of fully reconfigurable WPCNs follows the harvest-then-transmit protocol with integrated spatial reconfiguration, as illustrated in Fig. \ref{fig:workflow}. To support this operation, a central controller (e.g., at the BS) orchestrates the system-wide reconfiguration process. It collects CSI, energy-state information, and device status, and determines the configurations of PASSs, FASs, MAs, and RISs according to service requirements and channel conditions. A control link is responsible for delivering reconfiguration commands and collecting feedback, forming a closed-loop WET–WIT coordination mechanism that enables adaptation to dynamic environments and blockage conditions.

During the WET phase, the PASS-enabled PB transfers RF energy to IoT devices. The pinching points of PASS and the antenna ports of FAS are dynamically adjusted according to device locations and downlink channel conditions, enabling spatially focused and adaptive energy transfer. The harvested energy is stored at IoT devices and serves as the energy budget for subsequent uplink transmission.

During the WIT phase, IoT devices transmit information to the BS using the harvested energy. As illustrated in Fig. \ref{fig:workflow}, the FAS states at IoT devices, MA positions at the BS, and RIS reflection coefficients are jointly configured to enhance uplink channel quality and improve reception reliability, enabling device-aware and environment-aware adaptation of the propagation process.


Finally, CSI, energy status, and performance metrics (e.g., uplink throughput and outage probability) are fed back to the controller, forming a closed-loop adaptation process that enables continuous reconfiguration under dynamic channels, device distribution, and blockage conditions.

\begin{figure*}[!t]
	\centering
	\includegraphics[width=0.96 \textwidth]{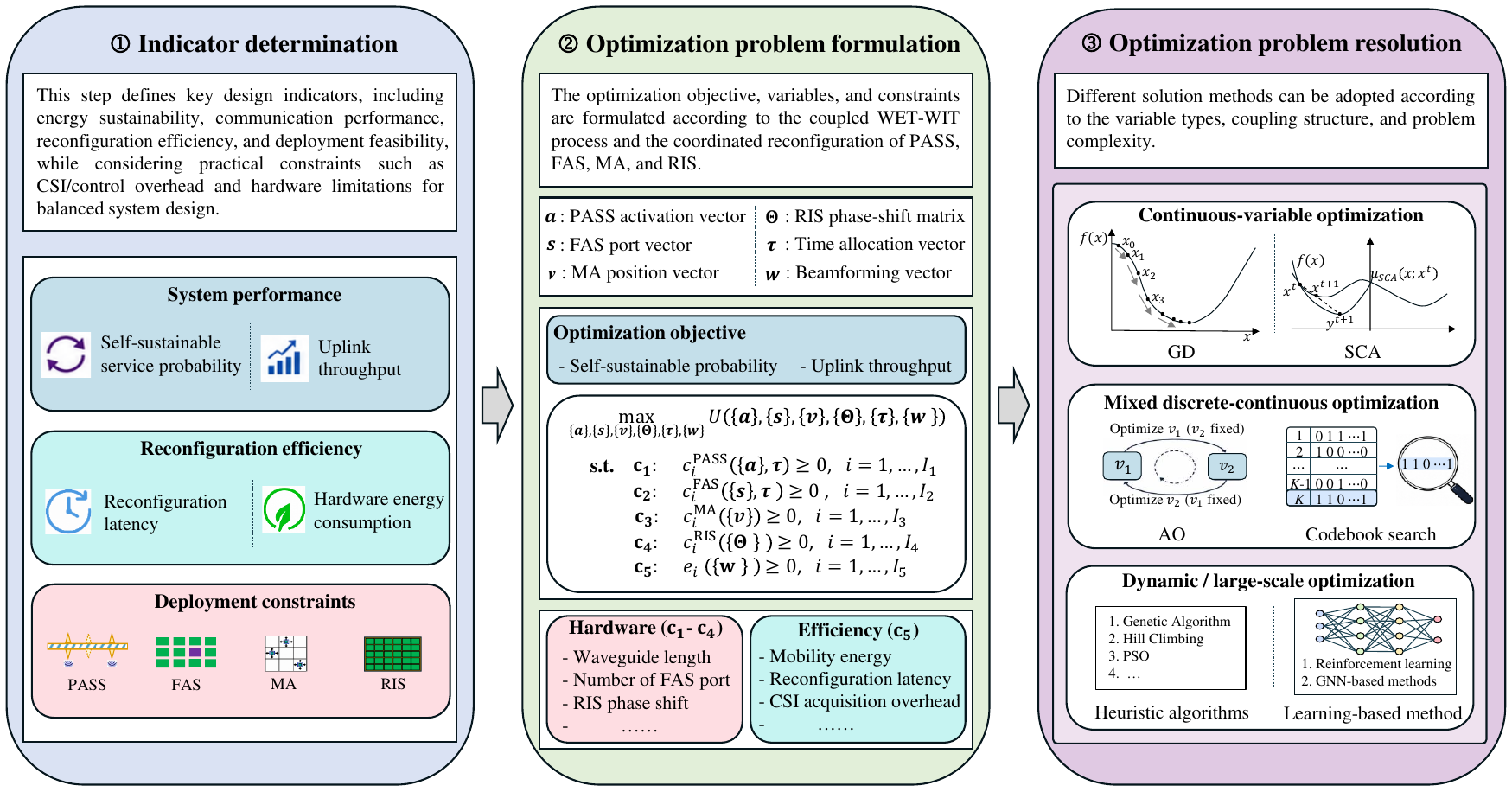}
	\caption{A general optimization framework for fully reconfigurable WPCNs.}
	\label{fig:Case2}
\end{figure*}

\section{Design and Optimization Framework}

Building on the unified architecture and closed-loop coordination process introduced above, Fig. \ref{fig:Case2} presents a general design and optimization framework for fully reconfigurable WPCNs. The framework follows a three-stage pipeline, including indicator determination, optimization problem formulation, and optimization problem resolution. It provides a unified way to organize the heterogeneous reconfiguration variables of PASSs, FASs, MAs, and RISs under the coupled WET-WIT process. It should be noted that this framework is not intended to define a single fixed optimization problem for all deployment scenarios. Instead, it provides a general modeling and solution guideline that can be adapted according to different performance objectives, hardware constraints, and application requirements. The mathematical form in Fig. \ref{fig:Case2} is intended to illustrate a generic optimization structure, while the specific utility function and constraints can be customized according to different deployment scenarios.

\subsection{Indicator Determination}

The first stage identifies the key performance indicators. Depending on the service requirement, the objective may focus on self-sustainable service probability, uplink throughput, or energy-information efficiency. Meanwhile, practical factors such as reconfiguration latency, hardware energy consumption, CSI acquisition overhead, and deployment feasibility should also be included. These indicators ensure that the system design reflects both communication performance and implementation cost.

\subsection{Optimization Problem Formulation}

The second stage formulates the optimization problem by specifying the variables, constraints, and objective function. The design variables may involve pinching-point activation, FAS port selection, MA positioning, RIS reflection control, WET/WIT time allocation, and uplink transmission strategies. The constraints are determined by hardware feasibility, finite candidate ports or positions, RIS phase control, energy causality, and delay requirements. Accordingly, the objective can be designed to maximize a weighted utility of self-sustainable service probability and uplink throughput.

\subsection{Optimization Problem Resolution}

The final stage selects suitable solution methods according to the variable types, coupling structure, and problem scale. Continuous variables, such as time allocation and beamforming, can be handled by gradient descent (GD) or successive convex approximation (SCA). Discrete variables, such as PASS activation, FAS port selection, and MA position selection, can be optimized by AO, codebook search, or other low-complexity heuristics. For dynamic and large-scale WPCNs, learning-based methods can further support environment-aware prediction and fast online reconfiguration. This framework therefore accommodates both model-based optimization and data-driven control for future fully reconfigurable WPCNs.

\section{Illustrative Performance Evaluation}

In this section, we conduct a representative numerical study for fully reconfigurable WPCNs. It should be emphasized that the following simulation is intended to provide an illustrative comparison of different WPCN architectures, rather than an exhaustive solution to the general optimization framework. Its purpose is to show the potential performance gain brought by coordinated spatial reconfiguration.

Following the architecture in Fig. 1, we establish the simulation setup summarized in Fig. 4 and evaluate the self-sustainable service probability and average uplink throughput. We compare four representative architectures, namely conventional WPCN, PASS-enabled WPCN, partially reconfigurable WPCN, and fully reconfigurable WPCN. The partially reconfigurable WPCN denotes the case where both the PB and IoT devices are equipped with reconfigurable antennas.

\begin{figure*}[!t]
	\centering
	\includegraphics[width=0.958 \textwidth]{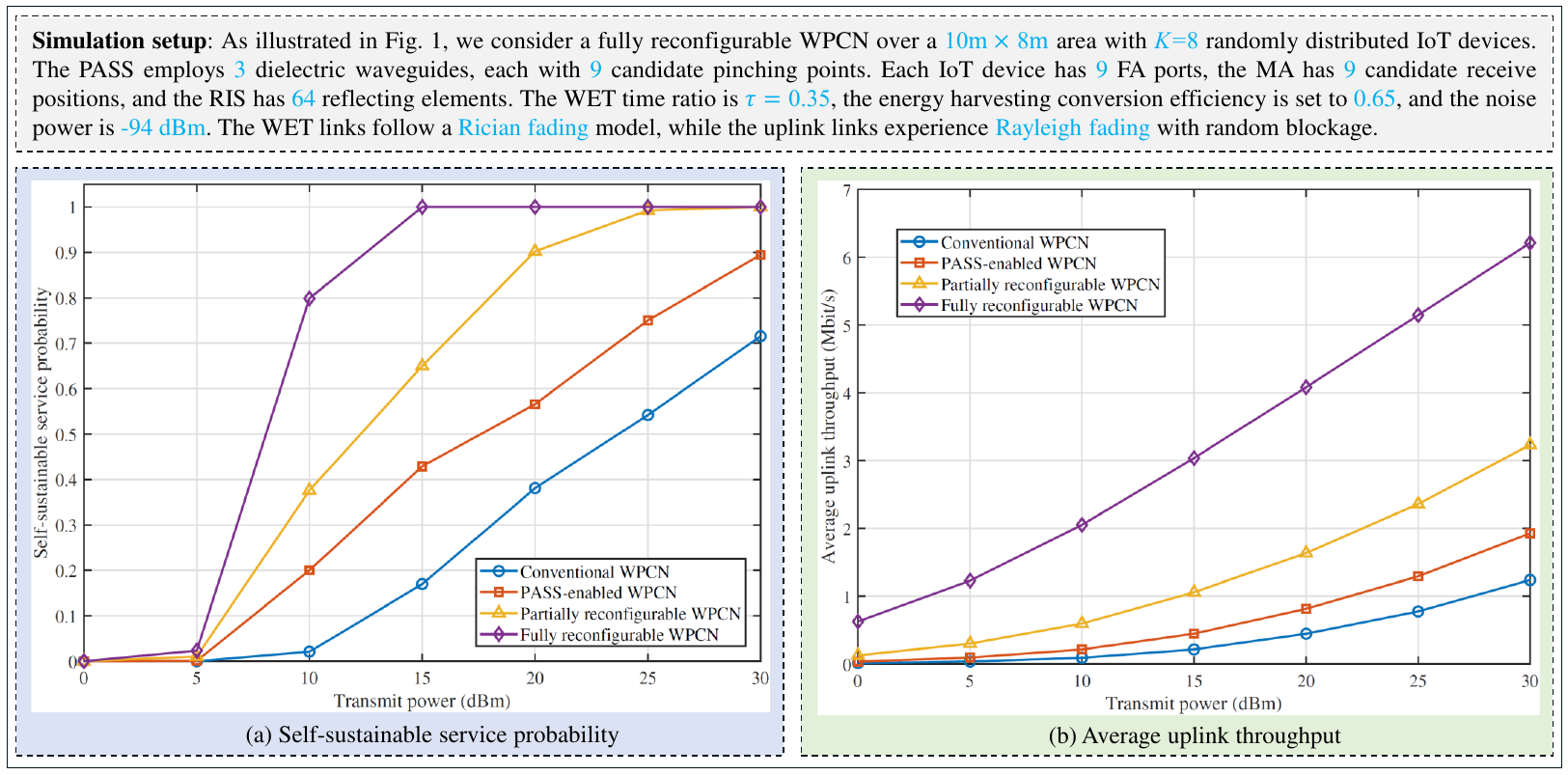}
	\caption{Simulation setup and performance comparison of different WPCN architectures.}
	\label{fig:Case1}
\end{figure*}

Fig. \ref{fig:Case1} compares the self-sustainable service probability and average uplink throughput under four architectures. The conventional WPCN yields the lowest performance, since both the energy-transfer link and the uplink information link are spatially fixed. By selecting favorable pinching points for WET, the PASS-enabled WPCN improves the harvested-energy budget and thus enhances the service probability and throughput. The partially reconfigurable WPCN further benefits from FAS-based device-side port selection. In contrast, the fully reconfigurable WPCN achieves the best performance by jointly enhancing WET through PASS/FAS and improving WIT through FAS, RIS, and MA reconfiguration. Moreover, its gain becomes more pronounced as the PB transmit power increases, because the additional harvested energy can be more effectively converted into uplink transmission capability and exploited by reconfigurable propagation and reception. These results show that improving WET or WIT alone is insufficient; instead, energy transfer, device-side adaptation, uplink propagation, and BS reception should be jointly coordinated from an end-to-end perspective.

It should be noted that the performance gains of fully reconfigurable WPCNs are accompanied by additional implementation costs. In general, incorporating more reconfigurable components can enhance the coupled energy-information transfer process, but it may also increase hardware cost, control signaling overhead, and configuration latency. These implementation-related costs are highly relevant in practical deployments and should be carefully balanced against the gains in self-sustainable service probability and uplink reliability. Therefore, fully reconfigurable WPCNs are most attractive when the achieved performance improvement can justify the increased system complexity. Table \ref{tab:tradeoff} provides a qualitative comparison of the implementation tradeoffs among different WPCN architectures.

\begin{table*}[!t]
	\renewcommand{\arraystretch}{1.15}
	\caption{Implementation Tradeoffs of Different WPCN Architectures}
	\label{tab:tradeoff}
	\centering
	\scriptsize
	\setlength{\tabcolsep}{2pt}
	\begin{tabular*}{0.78\textwidth}{@{\extracolsep{\fill}}p{0.22\textwidth}p{0.15\textwidth}p{0.19\textwidth}p{0.17\textwidth}}
		\toprule
		Architecture &
		Power/Circuit Cost &
		Reconfiguration Latency &
		Deployment Difficulty \\
		\midrule
		Conventional WPCN &
		Low &
		Low &
		Low \\
		PASS-enabled WPCN &
		Moderate &
		Moderate &
		Moderate \\
		Partially reconfigurable WPCN &
		Moderate to high &
		Moderate &
		Moderate \\
		Fully reconfigurable WPCN &
		High &
		High &
		High \\
		\bottomrule
	\end{tabular*}
\end{table*}

\section{Practical Challenges and Future Directions}

Although fully reconfigurable WPCNs provide a promising architecture for sustainable IoT networks, several practical issues must be addressed before they can be widely deployed. These issues arise not only from the limitations of individual reconfigurable components, but also from their joint operation within a coordinated WET-WIT system.

\subsection{Practical Challenges}

\subsubsection{Hardware prototyping and deployment cost}

Fully reconfigurable WPCNs require additional hardware compared with conventional WPCNs, including dielectric waveguides and pinching antennas for energy transfer, FAS structures at IoT devices, MA mechanisms at the BS, and RIS controllers for uplink propagation control. These components may increase installation cost, circuit complexity, mechanical maintenance, and energy consumption. Moreover, their practical performance may be affected by waveguide loss, antenna switching delay, MA movement accuracy, RIS control latency, and circuit power consumption. Therefore, prototype-driven evaluation is necessary to quantify the realistic performance-cost tradeoff.

\subsubsection{Runtime overhead and multi-timescale control}

Fully reconfigurable WPCNs require frequent interaction between sensing, decision-making, and reconfiguration control. However, different components operate at different time scales. RIS phase shifts and FAS states may be updated relatively frequently, while PASS pinching-point adjustment and MA movement may be slower due to mechanical or deployment constraints. In addition, collecting CSI and energy-state information from low-power IoT devices may introduce non-negligible signaling and energy overhead. How to coordinate heterogeneous reconfiguration actions without excessive runtime delay remains a key challenge.

\subsubsection{Reliability, security, and interoperability}

The operation of fully reconfigurable WPCNs depends on reliable control links and trustworthy feedback. Control errors, hardware failures, false energy-state reports, outdated CSI, or malicious attacks may degrade both WET and WIT performance. In addition, different reconfigurable components may be provided by different vendors and controlled through different interfaces, which makes system integration more difficult. Robust signaling, fault detection, secure control, and standardized interfaces are therefore essential for reliable and scalable deployment.

\subsection{Future Directions}

\subsubsection{Lightweight and hierarchical reconfiguration}

Future fully reconfigurable WPCNs should avoid optimizing all reconfigurable variables at the same time scale. A promising direction is hierarchical control, where slowly varying variables, such as PASS deployment and long-term RIS operating modes, are optimized at a coarse time scale, while FAS states, MA positions, and resource allocation are updated more frequently. Event-triggered reconfiguration can further reduce unnecessary updates by activating reconfiguration only when the channel, device distribution, or energy state changes significantly.

\subsubsection{AI- and digital-twin-assisted control}

AI techniques can help predict favorable energy-transfer regions, uplink transmission paths, and blockage patterns from historical CSI, sensing data, and device mobility information. Digital twins can further provide a virtual environment for evaluating different reconfiguration strategies before applying them to the physical network. Combining model-based optimization, learning-based prediction, and digital-twin-assisted validation may reduce online search complexity and support proactive reconfiguration in dynamic IoT scenarios.

\subsubsection{Prototype-driven evaluation and standardization}

Most existing studies on reconfigurable WPCNs are still based on theoretical models and simulations. Future research should develop small-scale prototypes to evaluate practical factors such as hardware power consumption, switching delay, control latency, and robustness under imperfect CSI. At the same time, standardized signaling protocols and control interfaces are needed for reporting CSI, energy states, device status, and reconfiguration commands. Such efforts will be important for moving fully reconfigurable WPCNs from conceptual architectures toward practical energy-sustainable IoT systems.

\section{Conclusion}
In this article, we have proposed a unified architecture for fully reconfigurable WPCNs by jointly integrating PASS-enabled PBs, FAS-based IoT devices, MA-assisted BSs, and RIS-aided propagation environments. By introducing end-to-end reconfigurability across WET and WIT, the proposed framework provides a new way to overcome the spatial limitations of conventional WPCNs. We have clarified the integration motivation and developed a general design and optimization framework covering indicator determination, optimization problem formulation, and optimization problem resolution. Illustrative performance evaluation further demonstrates the potential gain of coordinated spatial reconfiguration, while the implementation tradeoff analysis highlights the additional cost of full reconfigurability. Finally, we have discussed practical challenges and future directions related to hardware implementation, runtime overhead, reliability, security, and scalable control. This article provides new insights into the design of next-generation energy-sustainable IoT networks.

\end{document}